\documentclass[12pt,preprint]{aastex}
\usepackage{apjfonts,graphicx,graphics}
\usepackage{amsmath}
\usepackage{natbib}
\newcommand{\simless}
     {\ensuremath{\lower
3pt\hbox{$\rlap{\raise5pt\hbox{$\char'074$}}\mathchar"7218$}}}
\newcommand{\simgreat}
     {\ensuremath{\lower
3pt\hbox{$\rlap{\raise5pt\hbox{$\char'076$}}\mathchar"7218$}}}
\newcommand{\simgt}{\lower.5ex\hbox{$\; \buildrel > \over \sim \;$}}
\newcommand{\simlt}{\lower.5ex\hbox{$\; \buildrel < \over \sim \;$}}
\shorttitle{Magnetic Field Properties from Statistics}
\shortauthors{Koch et al.}
\begin{document}
\title{Magnetic Field Properties in High Mass Star Formation
from Large to Small Scales -
A Statistical Analysis from Polarization Data}
\author{
Patrick M. Koch\altaffilmark{1},
Ya-Wen Tang\altaffilmark{1} and 
 Paul T. P. Ho\altaffilmark{1,2}
}
\altaffiltext{1}{Academia Sinica, Institute of Astronomy and
 Astrophysics, Taiwan}
\altaffiltext{2}{Harvard-Smithsonian Center for Astrophysics, 60
 Garden Street, Cambridge, MA 02138, USA}

\email{pmkoch@asiaa.sinica.edu.tw}
%
%
\begin{abstract}

Polarization data from high mass star formation regions
(W51 e2/e8, Orion BN/KL) are
used to derive statistical properties of the plane of sky projected
magnetic field. Structure function and auto-correlation function
are calculated for observations with various resolutions from
the BIMA and SMA interferometers,
covering a range in physical scales from $\sim 70$~mpc
to $\sim 2.1$~mpc.
Results for the magnetic field turbulent
dispersion, its turbulent to mean field strength ratio and the 
large-scale 
polarization angle
correlation
length are presented as a function of the
physical scale at the star formation sites. Power law scaling relations
emerge for some of these physical quantities.
The turbulent to mean field strength ratio is found to be 
close to constant over the sampled observing range, with a hint of 
a decrease toward smaller scales, indicating that the 
role of magnetic field and turbulence is evolving with  
physical scale.
A statistical method is proposed to separate large and 
small scale correlations from an initial ensemble of polarization
segments. This also leads to a definition of a 
turbulent polarization angle correlation
length.

\end{abstract}
%
%
%
%
\section{Introduction}\label{intro}

Giant molecular clouds - the sites of star formation - are threaded
by magnetic fields. The exact role of the magnetic field, the nature
of turbulence and their interplay are still a matter of debate in
the literature. Evidence for a weak magnetic field (super-Alfv\'enic
turbulence) has been presented in e.g. \citet{crutcher09, padoan04},
whereas
support in favor of a strong magnetic field (sub-Alfv\'enic turbulence)
controlling the formation and evolution of the molecular cloud is
discussed in e.g. \citet{li09}. Accurate measurements of the
magnetic field strength are a key in distinguishing between these
two theories.

Different techniques have been employed to
measure the magnetic field intensity and structure on various scales.
The Zeeman effect provides the only known method of directly measuring
magnetic field strengths along the line of sight in a molecular cloud.
Generally, it has to rely on strong enough line intensities
and also high spectral resolution
in order
to detect the splitting of spectral lines
\citep{goodman89, crutcher99, troland08, crutcher09}.
Polarization of dust thermal emission at infrared and submillimeter
wavelengths provides another method to study magnetic field properties
\citep[e.g.,][]{hildebrand00}.
The dust grains are thought to be aligned with their shorter axis
parallel to the magnetic field lines, therefore the emitted light 
appears to be polarized perpendicular to the field lines
\citep{gk81, gk82, draine96, draine97, lazarian00}.
Radiative torques are likely to be responsible for the dust alignment
\citep{cho05, lazarian07}.
Complementary to Zeeman splitting, dust polarization measurements are
probing the plane of sky projected magnetic field direction. However,
in order to derive the actual magnetic field strength perpendicular
to the line of sight, additional assumptions are needed, as e.g. in
the commonly used Chandrasekhar-Fermi (CF) method \citep{chandra53}
or in its variations
\citep[e.g.,][]{houde04, curran07}.

When applying any of these CF methods, the dispersion in the measured
plane of sky polarization is a key parameter. Most studies up to now
relied on a magnetic field dispersion measured about a large scale
mean field
\citep[e.g.,][]{chrys94, lai01, lai02, tang09a}
or a model field
\citep[e.g.,][]{girart06,rao09}.
As noted in \citet{hildebrand09}, in dense clouds the magnetic field
structure might be the combined result from a variety of effects, such
as differential rotation, gravitational collapse and expanding HII
regions.
Consequently, a globally derived dispersion might not reflect the true
contribution from magnetohydrodynamic waves and/or turbulence. The
recent work by \citet{hildebrand09} develops a method based on a
dispersion function about local mean magnetic fields. Besides providing
a measure for the turbulent dispersion, the method also gives an
accurate estimate of the turbulent to mean magnetic field strength ratio.
Furthermore, the method is independent of any large scale field model.
\citet{hildebrand09} discuss applications to the Orion, M17 and DR21
molecular clouds, observed with the Hertz polarimeter at the Caltech
Submillimeter Observatory (CSO) with a resolution of 20$\arcsec$.

Dust polarization observations have been carried out over a range of
scales: from the large scale cloud envelope
\citep[e.g.,][]{schleuning98, lai01} to collapsing cores
\citep[e.g.,][]{girart06, tang09b}.
The goal of this paper is to apply and extend the method developed in
\citet{hildebrand09} across a range of scales in the same star formation
regions.

The paper is organized as follows: Section \ref{source} gives a brief
summary of the W51 e2/e8 and Orion BN/KL high mass star formation sites with the
results
relevant for our analysis. Section \ref{method} defines the structure
function and the auto-correlation function with some physical quantities
resulting from the statistical analysis. The results are presented
in section \ref{results} with a discussion in section \ref{discussion}.
Summary and conclusion are given in section \ref{summary}.

\section{Data Set and Source Descriptions} \label{source}

The presented data were obtained with the BIMA and SMA interferometers
at wavelengths where the polarization of dust thermal emission is traced.
The detailed descriptions of the data analysis and images are given in
\citet{lai01} for W51 e2/e8 with BIMA, \citet{tang09b} for W51 e2/e8 with
the SMA,
\citet{rao98}
for Orion BN/KL with BIMA and \citet{tang09c} for Orion BN/KL with the SMA.
Relevant observation numbers are given in Table \ref{quantities}.
The re-constructed features depend on the range of uv sampling
and weighting.
Nevertheless, these data currently provide the highest angular resolution
($\theta$) information
on the morphology of the magnetic field in the plane of sky obtained
with the emitted polarized light in those star formation
sites. They are thus complementary to the earlier analysis with single
dish data in \citet{hildebrand09} and references therein.
This study is part of the program on the SMA\footnote{
The Submillimeter Array is a joint project between the Smithsonian
Astrophysical Observatory and the Academia Sinica Institute of Astronomy
and Astrophysics, and is funded by the Smithsonian Institution and the
Academia Sinica.
}
 (Ho, Moran and Lo 2004) to study the structure of the magnetic field at
high spatial resolutions.

\subsection{W51 e2/e8}

W51 e2/e8 are some of the strongest mm/submm continuum sources in the W51
region.
Located at a distance of 7 kpc \citep{genzel81},
1$\arcsec$ is equivalent to $\sim$ 0.03 pc.
At a scale of 5 pc, measured at 850$\mu$m with a resolution 
$\theta\approx$ 15$\arcsec$ with SCUBA on JCMT \citep{chrys02, matthews09},
the polarization across the molecular cloud appears to be little organized 
and not uniform. However, the 350$\mu$m data, obtained with Hertz on the 
Caltech Submillimeter Observatory (CSO) show a well-ordered field
\citep{dotson10}. A comparison of the Hertz and SCUBA data is given in 
\citet{vaillancourt08a}. Both the e2 and e8 core are unresolved 
by Hertz and SCUBA.
\citet{lai01} reported a higher angular resolution (3$\arcsec$)
polarization map
at 1.3 mm obtained with BIMA, which resolved out large-scale structures.
In contrast to the larger scale polarization map,
the polarization appears to be 
more
uniform across the envelope at a scale of
0.5 pc,
enclosing the sources e2 and e8. In the highest angular resolution map
obtained
with the SMA at 870 $\mu$m with $\theta \approx 0\farcs$7,
the polarization patterns appear to be pinched in e2 and also possibly in
e8
\citep{tang09b}. 
In the following, where not explicitly written in full, W51 always 
refers to W51 e2/e8.

\subsection{Orion BN/KL}

The Orion Molecular Cloud (OMC-1) is one of the closest massive
star formation sites. At a distance of 480 pc \citep{genzel81},
1$\arcsec$ is equal to 2.3 mpc. Source BN/KL is located near the
strongest mm continuum emission, where the star formation process is
active.
Based on single dish polarization measurements obtained with SHARP on the
CSO, \citet{vaillancourt08b}
reported polarization maps observed at 350 and 450 $\mu$m in the OMC-1
ridge.
The revealed magnetic fields appear to be uniform across the dust ridge
with effective polarimetric beam sizes of 13$\arcsec$ at both
350 and 450 $\mu$m.
Similar uniform polarization maps have been reported at 100 $\mu$m by
\citet{schleuning98} and at 850 $\mu$m by \citet{vallee07} with
$\theta\approx 15\arcsec$.
Observed with BIMA at 1.3 mm and 3 mm with $\theta$ of $3\farcs4$
and 7$\arcsec$,
\citet{rao98} reported that the polarization appears to
change abruptly in the south of the mm continuum peak.
With the highest angular resolution achieved with the SMA at 870 $\mu$m,
$\theta\sim 1\arcsec$,
the polarization is consistent with the BIMA detections at 1.3 mm and 3
mm,
 but the field appears to vary smoothly across the entire core
\citep{tang09c}. 
In the following, where not explicitly written in full, Orion always 
refers to Orion BN/KL.

\section{Method} \label{method}

This section summarizes how the statistical quantities are derived from
the polarization data.

\subsection{Structure Function}   \label{structure}

The polarization position angle ($PA$) structure function (of second
order)
is a measure of the mean square deviation in the plane of sky projected
magnetic field directions\footnote{
Assuming the polarization emission to be perpendicular to the magnetic
field, the statistics derived from the $PA$ equally apply to the magnetic
field.
}
as a function of scale size.
Following the recent work by \citet{hildebrand09},
we adopt their definition for the magnetic field dispersion $\Delta \phi$,
which is the square root of the structure function:
\begin{equation}
<\Delta \phi ^2(l_k)>^{1/2}\equiv\left \{\frac{1}{N(l_k)}\sum_{l_k\le
r_{ij}<l_{k+1}}^{N(l_k)}
                           \left(\phi_i(r_i)-\phi_j(r_j)\right)^2\right
\}^{1/2},
                           \label{eq_structure}
\end{equation}
where $\phi_i=PA_i$ at position $r_i=(x_i,y_i)$ and
$r_{ij}=\sqrt{(x_i-x_j) ^2+(y_i-y_j)^2}$ in each source reference frame
with
coordinates $(x_i,y_i)$. $<...>$ denotes the averaging process over the
entire
polarization map with respect to $r_{ij}$ for each binning interval $l_k$.
$N(l_k)$ is the number of pairs of $PA$s with a separation in between
the bins $l_k$ and $l_{k+1}$.
All variables are plane of sky projected quantities.

As derived in \citet{hildebrand09}, assuming the magnetic field $B$ to be
composed of a smoothly varying large-scale
mean field $B_0$ and a statistically
independent turbulent
component $B_t$, the structure function in the range $\delta < l_k \ll d$
can be written as:
\begin{equation}
<\Delta \phi ^2(l_k)>_{tot}\simeq b^2+m^2 l^2_k + \sigma^2_M(l_k),
\label{fit_sf}
\end{equation}
where $\delta$ is the turbulent field correlation length and $d$ is the
typical scale for variations in $B_0$. $\sigma_M(l_k)$ are 
the binned error bars resulting from propagating the individual measurement 
uncertainties.
$b$ is interpreted as the scale-independent turbulent
field dispersion and $ml_k$ is the linear dispersion term (with slope $m$)
from the large-scale field $B_0$. All three contributions, being
statistically
independent, are added in quadrature.
This is basically a first order Taylor expansion of the structure function
in the domain where the turbulent field component, $B_t$, is a small
perturbation of a large-scale
field $B_0$ which is smooth on the scale of $d$.
Further following \citet{hildebrand09}, the ratio between the turbulent
and
mean magnetic field strength can be calculated by evaluating explicitly
the dispersion in the field directions under the assumptions of
small perturbations, which results in:
\begin{equation}
\frac{<B_t^2>^{1/2}}{B_0}=\frac{b}{\sqrt{2-b^2}}.     \label{ratio}
\end{equation}

\subsection{Auto-Correlation Function}

The polarization angle 
correlation function measures the resemblance
or self similarity
(correlation with itself)
of the projected polarization structure
on average as a function of separation.
Additionally, it leads to a definition of a characteristic polarization angle
correlation length. 
Whereas the ratio in equation (\ref{ratio}) relies on the assumption 
that the dispersion function can be written as a first order Taylor
expansion on the smallest scales, the auto-correlation function 
with its weighted moments (e.g. correlation length in 
equation (\ref{corr_length})) is independent of such assumptions.
In principle, this provides an independent cross-check 
(see section \ref{validity}) which can probe the assumptions in 
section \ref{structure}. Furthermore, higher resolution data 
will allow for an even more detailed modeling of the field
structure without relying on a fit.
We calculate the 
plane of sky projected polarization angle
correlation function
$\mathcal{C}$ as:
\begin{eqnarray}
<\mathcal{C}(l_k)>&\equiv&\frac{1}{N(l_k)}\sum_{l_k\le
r_{ij}<l_{k+1}}^{N(l_k)}
                   \phi_i(r_i)\cdot \phi_j(r_j) \nonumber \\
                  &\equiv& <\phi(r)\cdot \phi(r+l_k)>  \label{auto}
\end{eqnarray}
where the notation is identical to the one in equation (\ref{eq_structure}).
In order to make proper use of the auto-correlation function, one has to
assume homogeneous isotropic turbulence, i.e. $<\mathcal{C}(l_k)>$ must
depend only on the separations $r_{ij}$. This requires $\phi_i$ to be
rotationally invariant. Whereas this is naturally the case for the
structure
function, which involves only the square of the difference of the $PA$,
the product  $\phi_i(r_i)\cdot \phi_j(r_j)$ for the auto-correlation
function depends on the reference frame and the definition of the range of
the $PA$. The transformation $\phi_i \rightarrow \xi$,
$\phi_j \rightarrow \xi - \Delta \phi_{ij}$ expresses all the
correlation
products with respect to the same position angle $\xi(\neq 0)$,
where $\Delta \phi_{ij}=|\phi_i - \phi_j|$.
Since $0\le \Delta \phi_{ij}\le 90^{\circ}$, imposing zero correlation for
perpendicular
$PA$s fixes $\xi=90^{\circ}$. This definition also ensures the
correlation
coefficients to be in the range between zero and one ($PA$s parallel)
when normalized with $\xi^2$.

Each observed $PA$ will be the result of a superposition of a large-scale ($\phi_0$) and
a turbulent contribution ($\delta\phi$). Consequently, the correlation 
function as written in equation (\ref{auto}) contains both contributions
mixed. In order to characterize the large-scale polarization 
the turbulent part needs to be separated. Appendix A gives the details of 
how to derive a large-scale correlation function, $<\mathcal{C}_0(l_k)>$,
from an ensemble of measured position angles $\phi$, assuming
$\phi=\phi_0+\delta\phi$. 

The characteristic large-scale polarization angle
correlation length 
$\lambda_{0}$ 
is then calculated
by integrating the weighted large-scale polarization angle
correlation function:
\begin{equation}
\lambda_{0}=\frac{\int <\phi_0(r)\cdot \phi_0(r+l_k)>_r\cdot l_k \,\,dl_k}
               {\int <\phi_0(r)\cdot \phi_0(r+l_k)>_r \, dl_k}
\label{corr_length}
\end{equation}
where the integration extends over the entire binning range.
This is again a plane of sky projected quantity. In the case of a uniform
polarization, with all $PA$s being parallel and aligned with a single 
direction, all the correlation coefficients will be one, independent of 
scale. The large-scale 
correlation length 
{\bf
$\lambda_{0}$
}
 is then in the middle of
the largest and smallest scale, because all the scales are 
equally weighted (identical correlation) in this case.

We note that the correlation length, being an integrated and weighted
measure, is less sensitive to irregularities and incompleteness in the 
data set (see discussion in section \ref{validity}).
In analogy to 
equation (\ref{corr_length}), a turbulent polarization angle
correlation length
$\lambda_{t}$ can be estimated, once large-scale and turbulent
contributions are separated (Appendix A). 
A method to calculate the turbulent magnetic field correlation
length, based on a generalization of the dispersion function, 
is derived in \citet{houde09}.

\section{Results} \label{results}

Dispersion function and 
large-scale
auto-correlation function together 
with turbulent correlation function
are
presented in Figure \ref{w51_stat} and \ref{orion_stat}
for W51 e2/e8 and Orion BN/KL.
For both high mass star formation regions, the polarization data from
BIMA \citep{rao98, lai01} and the SMA \citep{tang09b, tang09c} are analyzed
following section \ref{method}. Table \ref{quantities} summarizes the
observations and our findings. The binning intervals $l_k$ are set to
integers of the synthesized beams.
For elliptical beams, resulting from a non-uniform $uv$-coverage, 
the geometrical mean is adopted.
Within each binning interval $k$, dispersion and auto-correlation
are evaluated for $l_k\le r_{ij}<l_{k+1}$.
Correlated data points below
the synthesized beam resolution are removed.
Only $PA$s with a polarized flux of more than 3$\sigma_{\rm I_{\rm P}}$,
the rms noise of the polarized intensity, are included.
Errors of individual $PA$s are typically in the range of $5^{\circ}$ to
$10^{\circ}$. The binned error bars 
($\sigma_M(l_k)$ in equation (\ref{fit_sf}))
in the Figures \ref{w51_stat} and
\ref{orion_stat} are then determined by propagating the individual errors
through the equations (\ref{eq_structure}) and (\ref{auto}).
For the dispersion function they are typically around $0\fdg5$ or less
for the smallest scales. This is due to the sample variance factor
from the large number of data points ($\sim$ 100 or more pairs of $PA$s).
They grow to a few degrees at the largest scales.
Fitting for the turbulent dispersion $b$ is based on a least square fit of
equation (\ref{fit_sf}).
The small binned error bars are neglected here, which is justified by
the possible larger biases as it is discussed in section \ref{validity}.

\subsection{W51 e2/e8}

\begin{figure}
 \begin{center}
 \includegraphics[scale=0.7]{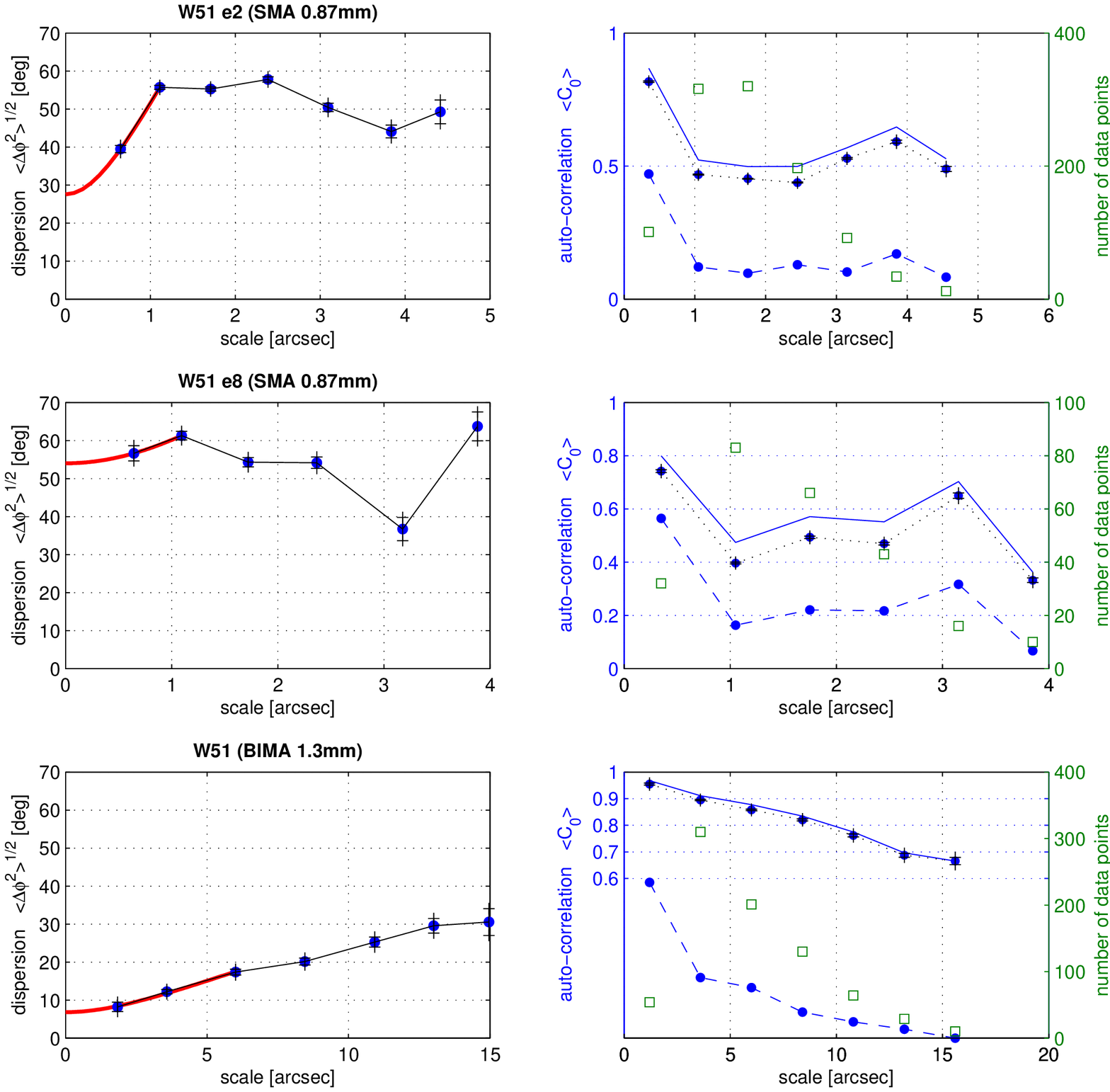}
 \caption{\label{w51_stat}\footnotesize
Left Panels: Dispersion (square root of second order structure function),
binned
to multiple integers of the resolution of each observation for W51 e2/e8. The
assigned
scale for each bin is obtained from averaging all the scales within a bin,
which is
very close to the synthesized beam in the first bin and 
close to the center for all the following bins.
Correlated data points below the synthesized beam resolution are 
removed.
Error bars are very small at the smallest scales due to the sample
variance
factor (large number of data points), and grow for the larger scales.
 The red line is the
fitting result including the first two or three bins following equation
(\ref{fit_sf}),
where the turbulent dispersion $b$ is found from the interception of the
fit
at scale=0.
Right Panels: large-scale
polarization angle correlation function $<\mathcal{C}_0>$ 
(solid line) binned
to multiple integers of the resolution of each observation. The
weighted integral of the curve measures the characteristic
large-scale polarization angle
correlation
length $\lambda_{0}$.
The number of data points (pairs of $PA$s) within each bin is displayed
($\Box$) with the axis on the
right hand side of each panel. 
The same number of data points are also used for calculating
the structure function, except where correlated data points
are removed. 
Binned values are connected with straight segments for visual guidance
only.
The dashed line shows the 
turbulent polarization angle correlation
function, separated
by the method described in Appendix A. The dotted line is the raw
auto-correlation function $<\mathcal{C}>$, directly derived from 
a measured ensemble of $PA$s without separating large and small scales.
At the distance of W51 ($\sim 7$~kpc), $1\arcsec$ corresponds to
$\sim 30$~mpc $\approx 6190$~AU.
 }
 \end{center}
\end{figure}

The statistical analysis for W51 e2/e8 was performed with three data sets:
BIMA \citep{lai01} at 1.3 mm with a resolution of $2\farcs3$,
covering the large-scale structure over $\sim 20^{\prime \prime}$,
and SMA at 870$\mu$m with a resolution of  $0\farcs7$,
separately resolving the regions e2 and e8 over about 4$^{\prime \prime}$
\citep{tang09b}. All dispersion functions (structure functions of
second order) show an increase over at least the first two bins
(Figure \ref{w51_stat}, left panels). The larger scale BIMA
measurement reveals a gentle increase in dispersion with a hint of
a plateau at the largest scales (Figure \ref{w51_stat}, left bottom
panel).
This is very similar to the results in \citet{hildebrand09}, obtained
with a 20$^{\prime \prime}$ 
resolution in M17, DR21 Main and
OMC-1. The higher resolution SMA observations show a steeper slope
over the first two bins, with a dispersion at the smallest scales of
about 40$^{\circ}$ and 55$^{\circ}$, compared to about 10$^{\circ}$ in the
BIMA
observation. Whereas increase at smaller scales and tendency of a
plateau at larger scales are still observed, the higher resolution
observations show more irregularities. This is particularly the case
for W51 e8.

Following \citet{hildebrand09} the turbulent field dispersion $b$,
as defined in
equation (\ref{fit_sf}), is obtained from the zero intercept of the fit
at scale$=0$. In order to stay in the linear regime, the first three
bins from the BIMA data and only the first two bins from the SMA data are
used.
(red solid lines in Figure \ref{w51_stat}, left panels). The resulting
turbulent dispersions around the mean local magnetic field range from
$\sim 6^{\circ}$ to $\sim 54^{\circ}$, with corresponding turbulent
to mean field strength ratios from 0.1 to 0.9 (equation (\ref{ratio})).
Higher resolution observations reveal larger values.

The large-scale polarization angle
correlation function for the BIMA observation shows a
smooth curve as expected from the dispersion
(Figure \ref{w51_stat}, bottom right panel, solid line): a small dispersion at
small scales translates into a close correlation at these scales.
At the BIMA largest scales, tracing the large-scale polarization
variations, the auto-correlation decreases accordingly. 
The SMA observations of both e2 and
e8 show correlations at the shortest scales which are followed by a
rather sharp drop and a plateau-like extension. 
This again reflects
the corresponding features in the dispersion functions. Both cores
show a secondary peak in the auto-correlation function at larger
scales, probably tracing symmetry features in the hourglass-like 
pinched field morphology.
Calculating the characteristic polarization angle
correlation length over the maximum
scale range, as introduced in equation (\ref{corr_length}), we find
$\lambda_{0}=230$~mpc for the BIMA observation and $\lambda_{0}=73$~mpc
and 63~mpc for e2 and e8, respectively.
Due to the relatively small field of view sampled in our observations, 
the values of $\lambda_{0}$ possibly represent lower limits\footnote{
Additionally, some information on the largest scales might be absent
due to the missing zero-spacing in the interferometric observations. 
However, most of the observations used here contain information from 
short baselines with lengths comparable to a few antenna diameters, 
and coherent large-scale structures are apparent in the polarization 
maps. Correlations on even larger scales -  also given the observed
trend of decreasing correlation coefficients with larger scales - 
are therefore likely to add only negligibly to $\lambda_{0}$.
}
.
This can 
then also explain why the correlation does not fully vanish 
at the largest scales.

The same panels in Figure \ref{w51_stat} also show the 
small-scale correlations, separated by 
the method described in Appendix A.
It is apparent that the 
turbulent correlation 
function, after an 
initial sharp drop, is still showing features similar to the 
large-scale function $<\mathcal{C}_0>$. This is a consequence of 
the weighting scheme outlined in Appendix A, where even at 
a larger scale small dispersion values can be accounted for 
a turbulent correlation
with a certain probability. Limiting
the turbulent correlation
 to within the first few bins yields
$\lambda_{t}$ between 25 and 45~mpc.

\subsection{Orion BN/KL}

\begin{figure}
 \begin{center}
 \includegraphics[scale=0.7]{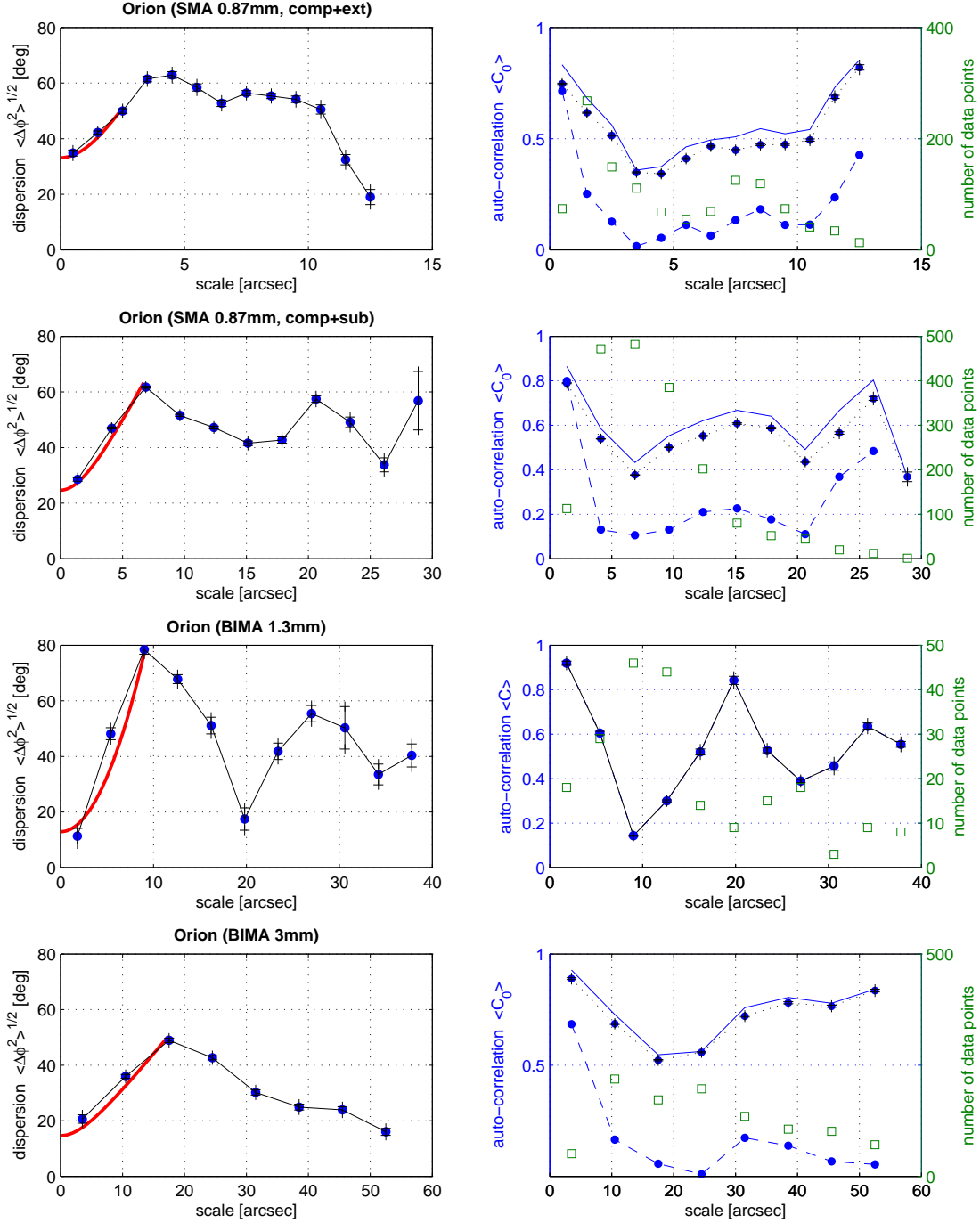}
 \caption{\label{orion_stat}
Identical to Figure \ref{w51_stat}, but for Orion BN/KL.
No satisfactory solution was found for Orion (BIMA 1.3mm).
The correlation shown is the raw polarization angle correlation 
function $<\mathcal{C}>$.
 At the distance of Orion ($\sim 480$~pc), $1\arcsec$ corresponds to
$\sim 2.3$~mpc $\approx 470$~AU.
}
 \end{center}
\end{figure}

Four data sets with very different resolutions were analyzed for Orion BN/KL:
BIMA observations at 3~mm and 1.3~mm with a resolution of 7$^{\prime
\prime}$
and  $3\farcs4$ \citep{rao98},
and SMA observations at 870$\mu$m from a combined compact with subcompact,
and a compact with extended configuration with resulting synthesized beams
of  $2\farcs8$ and  $0\farcs9$ \citep{tang09c}.
The general tendencies found for W51 e2/e8 - increase at smaller scales and
plateau-like extension with some irregularities at larger scales for the
dispersion function and their analogous features in the auto-correlation
function - are still present in Orion BN/KL (Figure \ref{orion_stat}).
However, already the lower resolution BIMA observations, in particular at
1.3~mm, show larger irregularities than the high resolution W51 e2/e8 data.
Since they are probing different physical scales at the corresponding
source distances (W51 being 14 times further away than Orion),
these larger irregularities might simply be due to
different morphologies and structures in the magnetic fields. In all
cases, the three lowest bins, showing a close to linear increase in
dispersion, were again used to fit for the turbulent dispersion $b$.
Values in the range of $\sim 13^{\circ}$ up to $\sim 33^{\circ}$ for
the highest resolution are found (Table \ref{quantities}). The
corresponding turbulent to mean field strength ratios are around 
0.2 
for
the BIMA and $\sim$0.4
for the SMA observations. Thus, over about
an eight times increase in linear resolution, dispersion and turbulent
field contribution increase by about a factor of three and two, respectively.

All the auto-correlation functions have in common a rather sharp drop
over the first three or four bins
(Figure \ref{orion_stat} right panels).
With successively higher resolutions,
this drop occurs within shorter scale ranges: within about $20^{\prime
\prime}$
and $10^{\prime \prime}$ for the BIMA observations, and $\sim 8^{\prime
\prime}$
and less than $5^{\prime \prime}$ for the SMA observations. In all cases
the auto-correlation function regains after this first drop, either with
a smooth slope (BIMA 3~mm, Figure \ref{orion_stat}, bottom right panel)
or with several peaks. The 
characteristic large-scale polarization angle correlation length,
equation (\ref{corr_length}), turns out to be in the range of $\sim
66$~mpc
to $\sim 15$~mpc, which is a change of about a factor of four.

The turbulent correlation function again shows a sharp drop over the first 
few bins. No satisfactory function was found for the BIMA 1.3 mm 
observation. This is likely due to the incompleteness of the  statistics 
(only few polarization segments) which the method in Appendix A 
relies upon. $\lambda_{t}$ derived from the first few bins
is between 36 and 9~mpc.

\section{Discussion}    \label{discussion}

\subsection{Validity and Robustness of Approach}  \label{validity}

Since the high resolution observations are revealing some differences
and likely probe a different
magnetic field regime (section \ref{physical}),
we re-assess here the validity of our approach
when deriving the turbulent dispersion and turbulent to mean field
strength
from equation (\ref{fit_sf}). In order to stay well within the linear
range, the derivation in \citet{hildebrand09} assumes $\delta< l_k \ll d$.
Whereas the turbulent magnetic field correlation length $\delta$ is
a fundamental limit related to the turbulence dissipation scale 
and the ambipolar diffusion scale $L$,
the upper limit $d$, the typical length scale for variations in the
large-scale mean field $B_0$, depends on the magnetic field under
consideration.
For the field threading a molecular cloud (envelope), $d$ will be on
the order of 100~mpc or more (e.g. envelope in W51 e2/e8). In the case of
a collapsing core, the remaining large-scale field is of the size
of the core diameter, which is around 50~mpc for W51 e2 \citep{tang09b}. 
It becomes
clear that the allowable interval $\delta< l_k \ll d$ shrinks with higher
resolution. Nevertheless, in such a cascading picture, a previously
small scale field component becomes the large-scale mean field
component at the next higher resolution. This still satisfies
the assumptions as long as the resolution is not too high.
It should be further stressed that the method in \citet{hildebrand09}
is entirely independent of a mean field modeling,
because of the restriction $ l_k \ll d$.
Some values of the derived 
turbulent polarization angle correlation
length $\lambda_{t}$
(Appendix A) are comparable to the smallest measured scales or
even larger. Therefore, $\lambda_{t}\sim\delta <l_k$ is, 
strictly speaking, not valid everywhere. Should the correlation length 
indeed be of the order of 10~mpc, some of our observations are already
resolving this scale. In such a case - referring to curve E in Figure 1
in \citet{hildebrand09} - the turbulence contribution to the dispersion 
is probably underestimated because the fitting is done in a range 
where the turbulence contribution has not yet reached its maximum.
The derived turbulent to mean magnetic field strength ratios are 
then lower limits.
We note that the highest resolution SMA observation in Orion
(Figure \ref{orion_stat} top left panel, $\theta \sim 2.1$~mpc)
is close to but not yet resolving the 
ambipolar diffusion scale $L$
($L\sim 1$~mpc from \citet{li08, lazarian04}).

In order to probe the robustness of the approach, a threshold test
is introduced. $PA$s with a continuum intensity above a certain
limit are excluded from the analysis. 
In the original work by \citet{hildebrand09} only data with a
magnetic field organization on large scales are analyzed. Our data
set additionally contains data where the magnetic field is 
organized on smaller scales. This is for example the case for the 
hourglass-like pinched
field lines in the collapsing core of W51 e2 where the central 
pinched field lines correlate with the strongest emission
\citep{tang09b}.
This threshold test addresses a possible
concern whether there is a bias toward the strongest emission data.
For Orion no relevant change in the dispersion function is found
until discarding $\sim 50$\% of the data ($\sim 70$\% cut in intensity).
In particular, the turbulent dispersion derived from the fit shows a
scatter within only $\sim \pm 2^{\circ}-3^{\circ}$, therefore not
altering the original results and proofing them to be quite robust.
This is verified for all the Orion data sets. Further excluding more
than 50\% of the data distorts the dispersion function, producing
unreliable fits and turbulent dispersion values. The same results hold
for W51 as observed with BIMA. The higher resolution observations of
W51 e2 and e8 however are found to be a little less robust.
For both e2 and e8 a decrease of
$\sim -7^{\circ}$ in the turbulent dispersion is found when excluding
up to 30\% of the data. Excluding more than 30\% of the data leads
to unreasonable results.
In the case of a
collapsing core (like e2 and possibly e8),
gravity pulls in the magnetic field lines and
presumably affects and dominates turbulence on some scales,
bending the field lines on the shortest scales possibly more than
what turbulence would do.
Removing the strongest intensity data -- which are in this case
closest to the center of the collapsing core -- is likely to reduce
a possible gravity induced bias on the shortest scales. The observed
decrease in the turbulent dispersion possibly reflects this.
In the subsequent Figure \ref{scaling_loglin}, these values are
shown with a down-arrow ($\downarrow$) indicating this possible bias.

Additionally, the dependence on the bin size has been checked.
Decreasing the bin size 
by about 30\%-50\%
of the synthesized beam leaves most of 
the results practically unchanged, with a typical
$\sim 1^{\circ}$ to $2^{\circ}$ down shift in the turbulent dispersion.
This results from the denser sampling at the shortest scales which
then extends the linear part of the dispersion function toward
a slightly shorter scale. This in turn leads to a slightly lower
value of the intercept at scale$=0$. Whereas oversampling shows
little effect, under-sampling with a twice as large synthesized
beam significantly biases the analysis toward larger dispersion
values. This can be understood from the steep slope over the first
few bins. Averaging over large scales then increases the dispersion
value. An exception to this conclusion are again W51 e2 and e8.
Oversampling by a factor of two reduces the turbulent dispersion
to $\sim 15^{\circ}$. This is possibly due to the same reason
as described above for the threshold test: Grouping the few values
with a large dispersion in a separate bin will lead to a much lower
dispersion for the remaining ones in another bin.
Orion, from the BIMA 1.3mm data, shows noticeable effects when the 
bin size is changed by only 10\% to 20\%. This is possibly due to 
the rather irregular dispersion function. 

In summary, the tests described above demonstrate that the analysis
gives generally robust results with a small scatter. The possible exception
is W51 e2 and e8, which might be biased toward too large dispersion
values. In the remaining sections, the discussion will be based
on the original values, with a reminder of the possible bias
where necessary.

The polarization angle correlation function $<\mathcal{C}(l_k)>$, being mathematically
related to the structure function via
$\Delta \phi ^2(l_k)\sim 2\left(<\mathcal{C}(0)>-<\mathcal{C}(l_k)>\right)$,
reflects the above discussed features analogously.
We remark that, although this mathematical connection exists, the
presented
auto-correlation functions,
 $<\mathcal{C}(l_k)>$ and  $<\mathcal{C}_0(l_k)>$,
 are calculated directly from the $PA$s
which provides thus an independent consistency test.
Mathematical relation and direct calculation are verified to lead
indeed to the same results
for $<\mathcal{C}(l_k)>$.

For the large-scale correlation function $<\mathcal{C}_0(l_k)>$, 
after separating the small scale turbulence contribution (appendix A), 
an immediate verification is not possible any more.
No reference in the literature
was found allowing a direct comparison here with other observations.
Technically, the relative errors for the auto-correlation function 
are smaller than for the dispersion function (less than 1\% compared 
to about 1 to a few percent), as a result of the error propagation
of the sum of products compared to the sum of differences.

Finally, we propose here to use the resulting 
large-scale polarization angle correlation
length $\lambda_{0}$ as a quantitative measure for the typical scale $d$ of
variation in the ordered polarization structure. The derived values (Table
\ref{quantities})
match reasonably well with the above quoted empirical values which are
estimated from the polarization maps. Our results are then based on the
dispersion function within $\delta < l_k \ll \lambda_{0}$.
Having a statistical measure for $d$ through the auto-correlation 
length $\lambda_{0}$ then also verifies the regime where the 
method based on the dispersion function (section \ref{structure})
can be applied.
Since $ \lambda_{0}$ is an integrated and weighted measure, threshold test
and bin size show only negligible changes. In the course of the 
tests described above, 
oversampling and under-sampling by a factor of two give 
variations in $\lambda_{0}$ of  
less than 10\% for all observations.

\subsection{Comparison with Previous Results}

Generally, the dispersion functions derived from the present
observation set for both W51 e2/e8 and Orion BN/KL are less smooth 
than those derived
from lower resolution single dish observations \citep{hildebrand09,
kirby09}.
We speculate that these irregularities are tracing underlying changes
and structures in the magnetic field morphology which become more
manifest only at higher resolution ($\theta$), but are smoothed out at
lower
resolution.
The previous larger scale (lower resolution) single dish observations show
polarization patterns which extend to such radii where the
magnetic field straightens out as the gravitational influence
is weak (e.g. OMC-1 in \citet{schleuning98} with the
Kuiper Airborne Observatory (KAO) at 100$\mu$m with a resolution of
$\sim 35 \arcsec$ and in \citet{houde04} with
the Hertz polarimeter at CSO with
$\theta \sim$20$\arcsec$ at 350$\mu$m).
M17 SW, observed over a $\sim 6^{\prime} \times 6^{\prime}$
field with the KAO shows an overall orderly field
configuration with a hint of being pulled into the cloud core by
gravitational collapse \citep{dotson96}. At the smallest separations,
the average change in polarization is around 10$^{\circ}$. The sample
of 12 Galactic clouds observed with the KAO shows mostly organized
large scale fields \citep{dotson2000}. No dispersion functions were
derived for this sample.

Signs of a collapsing cloud in the center, with otherwise mostly
straightened field lines, are observed for DR21 Main \citep{kirby09}.
A smooth dispersion function is found with a dispersion of $\sim
10^{\circ}$
in the lowest bin and a linear increase to $\sim 25^{\circ}$ over
the next three bins.
The one case in which we observe 
an equally smooth
dispersion is W51 observed with BIMA. Similarly to DR21 Main, this
observation is tracing the large-scale field envelope\footnote{
The smaller scale field structures are likely to be beam-diluted
because of averaging in the plane of sky, but not along the line
of sight. In the opposite case, higher angular resolution data,
e.g with the SMA, would not be able to reveal coherent structures on
smaller scales.
}
, but without
yet revealing a collapsing core \citep{lai01}. The lowest bin dispersion
is again around $10^{\circ}$, then linearly increasing over the
next four bins to about $30^{\circ}$ (Figure \ref{w51_stat} bottom
left panel). The single dish (DR21 Main with CSO, spatial resolution
$\sim 0.2$~pc) and interferometer
(W51 with BIMA, spatial resolution $\sim 0.1$~pc) seem to be revealing
comparable structures and
dispersion values around 10$^{\circ}$ here.
Such dispersion values then lead to a turbulent to mean field strength
ratio of 
0.1 to 0.15. 
These numbers are also consistent
with the values reported recently in \citet{hildebrand09} for OMC-1
and M17.
Our remaining data are directly probing the collapsing cores with
clearly pinched magnetic field lines for W51 (see Figure 6 in
\citet{tang09b})
and a wrapped toroidal-like structure in Orion \citep{tang09c}.
For this different physical
regime - although previous observations exist \citep{girart06} - no
analysis based on a dispersion function was performed which would
allow a comparison. A turbulent to magnetic energy ratio of $\sim 0.02$
(turbulent to magnetic field strength ratio $\sim 0.14$) is derived by
\citet{girart06} based on a remaining mean dispersion after fitting
a parabolic function to the field morphology.

A further major difference between our data set (except W51 with BIMA)
and the above cited previous works is the steeper slope
(section \ref{numerical}, Table \ref{quantities})
over the
first two or three bins. The dispersion increases typically by
20$^{\circ}$
to 30$^{\circ}$ or even more, compared to $\sim 10^{\circ}$ in the
above cited cases.
The values at the first bins are already around
20$^{\circ}$ or 30$^{\circ}$ whereas the other cases show numbers around
10$^{\circ}$ or less. The corresponding turbulent to mean field strength
ratios are then in the range of $\sim$ 
0.2 to 0.54.

Different methods, also aiming at constraining the turbulent to
mean field ratio but not going through a structure function, have
been explored by several authors. Based on an average $PA$
dispersion $\delta \phi < 13^{\circ}$, a turbulent to magnetic
field energy ratio has been estimated in \citet{lai02}
for NGC 2024 FIR 5 in the Orion B Giant Molecular cloud. Their
energy ratio (<0.14) corresponds to a turbulent to mean field strength
ratio <0.37.
The ratio of mass-to-flux, $M/\Phi$, is evaluated in \citet{troland08}
for a set of 34 dark cloud cores ($\sim 0.01$~pc) from OH Zeeman
observations with the Arecibo telescope. Their average ratio of
turbulent to magnetic energy is $\sim$2, the turbulent to mean field
ratio therefore about 1.4. Part of these data, in combination with
GBT observations of the cloud envelope ($\sim$ 1~pc), were then
used to provide support for a super-Alfv\'enic (weak magnetic field)
turbulence model \citep{crutcher09}.
Average dispersion values
(not dispersion functions) around a mean magnetic field direction
were also derived in \citet{myers91} from optical polarimetry and
in \citet{novak09} from submillimeter polarization.
In the latter work the authors found a turbulent to mean field
ratio in the range between 2.0 (intermediate field) and 0.52
(strong field) compatible with their observations.

\subsection{Comparison with Numerical Simulations} \label{numerical}

The turbulent to magnetic field energy ratio is also investigated
through numerical simulations. \citet{ostriker01} analyzed the time
evolution in a model cloud simulation with three different magnetic
field strengths (strong, medium and weak field with
$\beta=c_s^2/B_0/(4\pi \bar{\rho})=0.01$, 0.1 and 1, respectively),
but an identical initial turbulent velocity field. The perturbed
magnetic field energy reaches a maximum at about 0.1 - 0.2 times the
Alfv\'enic crossing where it accounts for about 20 - 50\% of the total
turbulent energy (kinetic energy and perturbed magnetic energy together).
In their projected snapshot only the strong field model ($B_0=14~\mu$G)
leaves significantly correlated ordered polarization segments with a
perturbed to mean magnetic energy $\beta_{turb}=\delta B^2/B_0^2\approx
0.27$.
This corresponds to a turbulent to mean field strength ratio of about 0.52.
Although this is comparable to the numbers of our higher resolution
observations around collapsing cores, it is not obvious to match their
snapshot in time evolution within our sequence of low and high resolution
data.

On the other hand, recent simulation results with supersonic and
super-Alfv\'enic turbulence,
in combination with simulated Zeeman measurements, find
$\beta_{turb}\simgt 1$
for cores with a radius $\sim 0.2 - 0.8$~pc \citep{lunttila09}.
This seems to be in favor of
the turbulent to magnetic energy ratios observed in \citet{troland08},
who found an average $\bar{\beta}_{turb}\approx 2$ for comparable core
sizes and source distances. These findings support a star formation
theory with super-Alfv\'enic turbulence.

Related to $\beta_{turb}$, $PA$ structure functions (of second order),
$\Delta \phi^2$,
are analyzed in \citet{falceta08} with the goal of discriminating
between sub/supersonic and sub/super-Alfv\'enic models. Independent
of the magnetic field orientation with respect to the line of sight,
sub-Alfv\'enic models tend to have a power law index $\alpha \sim 0.5$,
$\Delta \phi^2(l_k) \sim l_k^{\alpha}$, whereas super-Alfv\'enic
models show a flatter slope with $\alpha \sim 0.3$. For the set of
our observations, $\alpha$ (derived from the first two bins for W51
and three bins for Orion, Table \ref{quantities})
is in the range of 0.3 to 3.2. Although this is closer to
sub-Alfv\'enic models, the rather steep slopes in our
structure functions followed by a plateau make a direct comparison
with the smoother structure functions in \citet{falceta08} not
obvious.

In summary, numerical simulations provide results for the turbulent
to magnetic field energy ratio and the slope of the polarization
structure function for a series of different models. Some are in
agreement with our findings. For a detailed comparison, the
difficulty lies in matching observational resolution and star
formation stage with the model time evolution in the simulation.

\subsection{Dependence on Physical Scale} \label{physical}

The angular resolutions obtained from the BIMA and SMA observations 
vary by about a factor
of ten, spanning a range in physical scales at the source distances
from $\sim 70$~mpc to $\sim 2.1$~mpc (Table \ref{quantities}).
As discussed in section \ref{results}, the resulting dispersion
functions (and auto-correlation functions) show some common
tendencies but differ in their detailed characteristics and numbers.
Here, we address the question whether the subsequently higher
resolution observations 
reveal a dependence on physical scale.
This question is also motivated 
by the observed polarization maps
where an envelope on the largest scales (BIMA, \citet{lai01}) and
two collapsing cores with a possible hourglass-like magnetic field
morphology on the smallest scales 
(SMA, Figure 6 in \citet{tang09b}) are found in the case of W51 e2/e8.
For Orion, the lowest resolution BIMA data
and the highest
resolution SMA data again reveal very different magnetic field 
structures.
Consequently, we argue that
these polarization observations are indeed
probing different regimes in the star formation process.
The fact that increasingly higher resolutions still reveal
detectable local coherent structures, very likely means that 
large-scale structures are smooth enough for the local structures
to be distinguishable. This must hold true for both along and 
across the line of sight, because otherwise features would be
washed out.

\begin{figure}
 \begin{center}
 \includegraphics[scale=0.7]{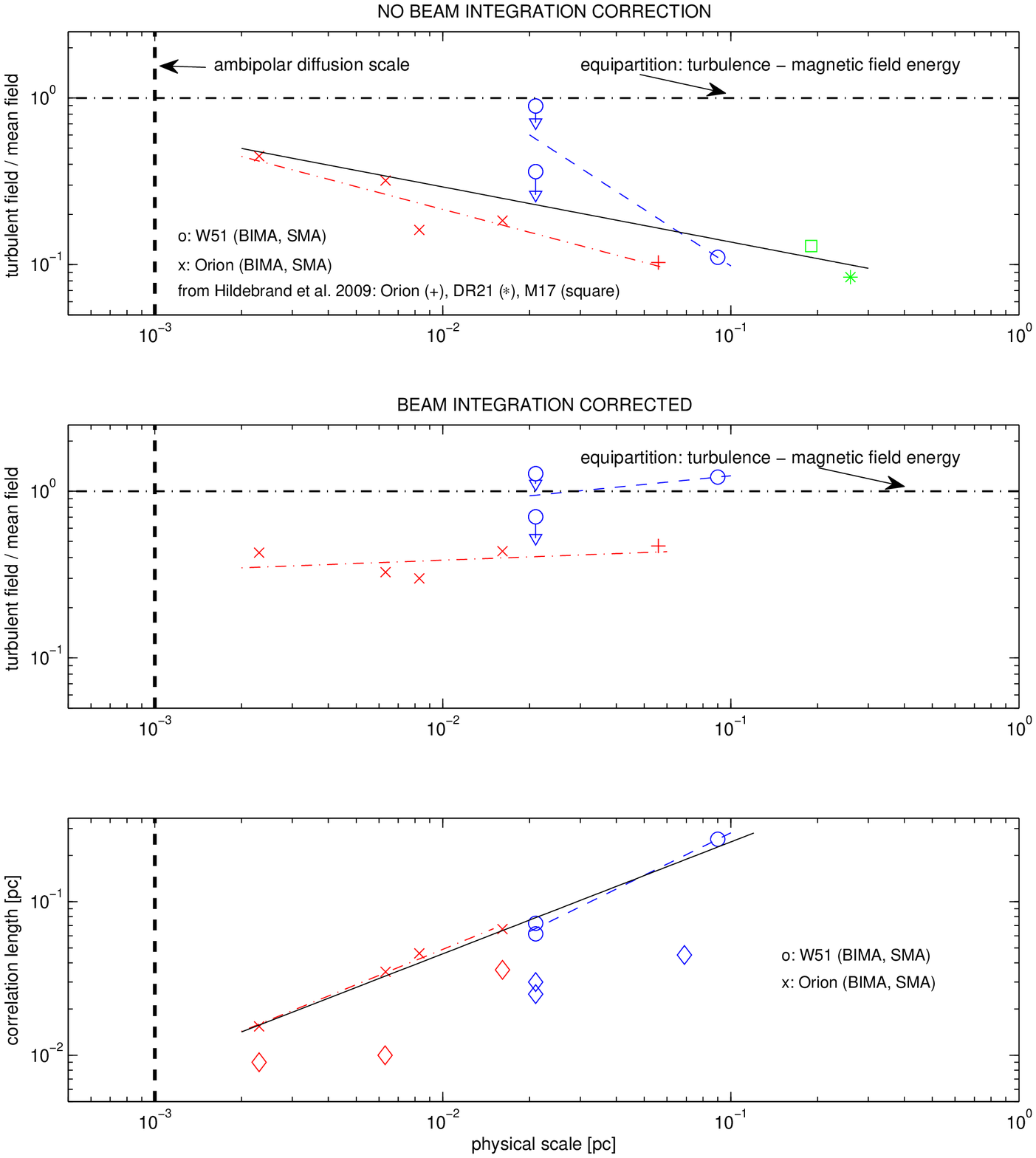}
 \caption{\label{scaling_loglin} \footnotesize
  Top panel: turbulent to mean field strength ratio
$<B_t^2>^{1/2}/B_0$ from various observations as a function of
physical scale, as derived from the Figure \ref{w51_stat} and 
\ref{orion_stat}.
 As relevant physical scale the achieved resolution
(synthesized beam) at the source distance is assigned.
Added to the BIMA and SMA data set are
DR21 ($\ast$),
M17 ($\Box$) and Orion ($+$) from \citet{hildebrand09}.
The down arrows for W51 e8 and e2 mark the values after correcting for a
possible
gravity induced bias ($\sim -7^{\circ}$) as discussed in section
\ref{validity}.
The middle panel shows the same ratio, but 
taking into account a beam integration correction as described in 
section \ref{physical}.
Resulting power law fits $(\propto l^{\gamma})$ in both panels
are shown separately for the original W51 data (blue, dashed line), Orion
(red, dot-dashed line) and the
entire data set (black solid line, only top and bottom panel). The indexes are summarized in Table
\ref{power}.
For illustration shown is also the expected
ambipolar diffusion scale at $\sim 1$~mpc from \citet{li08} (dashed vertical line)
and the turbulent - magnetic field strength equipartition line 
(dot-dashed line).
Bottom panel: large-scale polarization angle correlation length, 
$\lambda_{0}$, as a function of physical scale, together 
with the 
turbulent polarization angle correlation
lengths (diamonds).
 }
 \end{center}
\end{figure}

Figure \ref{scaling_loglin}, top panel, shows the tendencies
for the turbulent to mean field strength ratio as a function 
of the physical scales, as derived
from
the  Figures \ref{w51_stat} and \ref{orion_stat}. As relevant physical
scale
the synthesized beam resolution is assumed.
In addition to the BIMA and SMA data set, the lower resolution data from
\citet{hildebrand09} for Orion, DR21 and M17 are also added.
Over the observed scale range ($\sim 70$~mpc to $\sim 2.1$~mpc),
the turbulent to mean field strength ratio increases with 
smaller scale by about a factor of ten.

Further investigating this apparent dependence on physical scale
in the top panel in Figure \ref{scaling_loglin}, we discuss 
a possible beam resolution effect. \citet{houde09}, 
in an expansion on the work in \citet{hildebrand09}, considered
the signal integration aspect in their analysis. They find that 
the turbulent component of the dispersion function at scale zero, 
$b^2(0)$, is then the square of the turbulent
to large-scale magnetic field strength divided by the number of 
independent turbulent cells $N$ probed by the observation:
$b^2(0)=<B_t^2>/B_0^2\cdot 1/N$. As further derived, $N$ is directly
a function of the beam size and can be written as 
$N=\frac{(\delta^2+2W^2)\Delta^{\prime}}{\sqrt{2\pi}\delta^3}$, 
where $\delta$ is the turbulent field correlation length, 
$\Delta^{\prime}$ is the effective depth of the molecular cloud
along the line of sight and $W$ is the beam radius. 
The equation for $N$ in \citet{houde09} is derived 
assuming a (circular) Gaussian beam and Gaussian turbulent
auto-correlation functions.
For a given 
source, the smaller the beam size, the smaller $N$ will be, 
and the bigger $b$ will be if the ratio $<B_t^2>^{1/2}/B_0$
remains constant, or if at least $1/\sqrt{N}$ grows faster than
$<B_t^2>^{1/2}/B_0$ decreases.
Different beam sizes can therefore mimic a trend in the 
turbulent to mean field ratio.  
The top panel in Figure \ref{scaling_loglin} apparently confirms
this expectation.
In the following we aim at revising the top panel in  
Figure \ref{scaling_loglin} by taking into account the beam
resolution effect, in order to reveal the net change in 
the turbulence to mean field strength ratio over scale.
Starting from equation (44) in \citet{houde09}, the scale
independent term describes the intercept at scale zero in their
Figure 2. This is equivalent to our dispersion at scale zero,
but with an additional factor containing the beam integration
effect. We, therefore, rewrite the turbulent to mean magnetic
field ratio as: $<B_t^2>^{1/2}/B_0=\sqrt{N}/\sqrt{2}\cdot 
<\Delta \phi^2>^{1/2}_{l=0}$, where $<\Delta \phi^2>^{1/2}_{l=0}$
is the turbulent dispersion derived from our fitting results
in Figure \ref{w51_stat} and \ref{orion_stat}. 
$\Delta^{\prime}$ is approximated with the size of the most
extended detected structure, which is roughly the size of the 
maps in the SMA and BIMA observations. $\delta(\approx$ 16~mpc)
is adopted from \citet{houde09} for Orion. The beam radius
$W$ is derived from the synthesized beams of our 
SMA and BIMA observations. The estimated number of turbulent 
cells is listed in Table \ref{quantities}. Figure \ref{scaling_loglin}, 
middle panel, shows the resulting beam corrected turbulent to 
mean magnetic field strength ratios. After correction, both 
Orion and W51 show a close to constant ratio over scale, with 
a slight trend of a decreasing ratio toward smaller scales. 
Whereas the ratios for Orion are in the range between 0.30 and 
0.44, the ratios for W51 (0.7 to 1.27) are around the equipartition limit
of turbulent and magnetic field strength. This might be an 
indication that the role of turbulence and magnetic field changes
over scale. In the case of W51, there is possibly a transition from a 
turbulence dominated to a magnetic field dominated scenario.

Taking into account the beam integration modifies the picture 
first presented in the top panel.
Differences in measured values of 
$<\Delta \phi^2>^{1/2}_{l=0}$ can result from a combination of a 
beam integration effect and a variable turbulent to mean field ratio.
Both contributions can vary in their significance with scale.
Correcting for the beam integration effect reveals
the net change over scale in the turbulent to mean field ratio.
The beam correction ($\sim\sqrt{N}$) is of importance for lower resolution 
(larger beam) observations, but becomes less important for 
higher resolution observations approaching the turbulent cell size.
Intuitively, one might expect a turbulent to mean field strength
ratio which decreases with smaller scale as the turbulence 
dissipation scale is approached. 
The observed trend seems to be in support of this, revealing the 
beginning of a slight decrease in the ratio. 
An even higher resolution 
observation at the mpc scale or smaller might then
reveal a breakpoint (turn over) in the scaling. On the other hand, 
the panels in Figure \ref{scaling_loglin} only display a ratio.
It is still possible that the turbulent field strength 
increases with smaller scale, but at a slower rate than the 
mean field strength.  
A power law,
$\propto l^{\gamma}$, is fit to the Orion and W51 data. 
Table \ref{power} summarizes
the results
for both the uncorrected and the beam corrected scalings. 

We finally remark that there is some uncertainty left
in the beam correction due to our approximation for $\Delta^{\prime}$.
Should the detected cores have structures along the line of sight
which significantly differ from the detected extension across the 
plane of sky, the number of turbulent cells $N$ could vary. 
Underestimating $N$ by a factor of two would change the ratio by 
$\sqrt{2}$. In order to align Orion and W51, $N$ would need to be 
underestimated or overestimated, respectively, by about a factor 
of four. Similarly, a correlation length $\delta$ twice as large
as in Orion would reduce the ratios for W51 to approximately the 
ones found for Orion. However, remaining uncertainties in both 
$\Delta^{\prime}$ and $\delta$ do not change the slope of the 
scaling, unless they significantly vary with resolution.
Thus, within these uncertainties Orion and W51 both show indications
of a decrease in the turbulent to mean field strength ratio over scale, 
but with a difference in scale specific ratios possibly reflecting 
the source environment.

The scaling
of the large-scale 
polarization angle correlation length $\lambda_{0}$ shows a close to
straight line (Figure \ref{scaling_loglin}, bottom panel),
with only a small difference between W51 and Orion.
$\lambda_{0}$ decreases by about a factor of fifteen over the sampled physical
scales.
We remark that the presented values for $\lambda_{0}$ are possibly
lower limits, because of the relatively small 
fields of view
sampled in 
our observations. 
Since $\lambda_{0}$ is a normalized measure, equation (\ref{corr_length}),
beam integration effects are likely to cancel out, unless they 
additionally depend on the spatial scale $l_k$.
The turbulent correlation length $\lambda_{t}$ is presumably
independent of scale. This is not fully verified from our data set
(Figure \ref{scaling_loglin}, bottom panel). In some cases 
this is due to incomplete turbulence statistics, which then do not 
allow to set a reliable cut off criteria. Additionally, a certain 
beam integration effect is probably left in $\mathcal{D}(\Delta \phi_{k=1})$, 
equation (\ref{turb_dist}), which leads to broader or narrower 
turbulence distributions. This again affects the cut off criteria
through $\mu_{\Delta}$ (Appendix A).  
For comparison, a 
turbulent magnetic field correlation 
length of $\approx$ 16~mpc is derived 
in \citet{houde09} for OMC-1.

\section{Summary and Conclusion}  \label{summary}

A set of interferometric polarization observations with resolutions
in the range between $7\arcsec$ and $0.\arcsec 7$
($\sim $70~mpc to $\sim$ 2.1~mpc)
is analyzed to derive statistical properties of magnetic field and
turbulence from large to small scales during the star formation process.
Our data set covers structures from the large-scale cloud envelope
to the collapsing cores. The highest resolution data are close to the
expected ambipolar diffusion scale
($\sim 1$~mpc). We apply and expand the method
developed recently in \citet{hildebrand09}. The main results are:

\begin{enumerate}

\item
The turbulent field dispersion shows a steeper slope and larger values
at the shortest scales with increasing resolution. 
Accordingly, the resulting turbulent
to mean magnetic field strength ratio increases with smaller scale 
over the entire range in physical resolution.
This is without taking into account a beam integration aspect, and 
is in agreement with earlier theoretical expectations.

\item
The sequence of low and high resolution observations  
does not only zoom
in onto the same magnetic field structure, but it is probing
different morphologies and different stages in the star formation process. This is also supported
by the polarization maps. 
When taking into account a beam integration aspect, 
both 
Orion and W51 show a close to constant turbulent to mean field strength
ratio over scale, with 
a slight trend of a decreasing ratio toward smaller scales. 
Whereas the ratios for Orion are in the range between 0.30 and 
0.44, the ratios for W51 (0.7 to 1.27) are around the equipartition limit
of turbulent and magnetic field strength. This might be an 
indication that the role of turbulence and magnetic field changes
over scale. In the case of W51, there is possibly a transition from a 
turbulence dominated to a magnetic field dominated scenario.
Our observation set therefore also provides
information for the time and spatial evolution of these 
quantities during the
star formation process.

\item
Based on the polarization angle correlation function a 
characteristic large-scale
correlation length
$\lambda_{0}$ is defined. This can be used as a quantitative criterion to
define the scale over which the mean polarization structure varies. 
$\lambda_{0}$ decreases with higher resolution.
Additionally, starting from an ensemble of measured polarization 
position angles, a method is proposed to separate statistically 
large-scale from turbulent contributions. This leads to a 
definition of a turbulent correlation length $\lambda_{t}$.

\end{enumerate}


The authors wish to thank the referees, 
Roger H. Hildebrand, Martin Houde and John E. Vaillancourt
for their comments and explanations
which provided important further insight.

\section*{A. Polarization Angle Correlation Function}

An observed position angle $PA_i=\phi_i$ is the superposition of a 
large-scale polarization structure and a smaller scale turbulent component. 
Consequently, 
a correlation length directly derived from a measured ensemble of $\phi_i$
contains both small scale and large-scale correlations mixed. In order to 
calculate a large-scale (mean) polarization angle correlation length separately, one 
needs to isolate the mean contribution $\phi_{i0}$ and the turbulent
contribution $\delta\phi_i$. In general, this will not be possible for 
a single or only a few $PA$s, unless one makes very specific assumptions
as, e.g. a given uniform magnetic field direction. However, it is, as 
outlined in the following, possible to separate the two contributions in 
a statistical way. 

Splitting each $PA_i$ into a large-scale and turbulent part, 
$\phi_i=\phi_{i0}+\delta \phi_i$, the correlation product for a scale $k$
is written as
\begin{equation}
\sum \phi_i\cdot \phi_j \sim\sum \phi_{i0}\cdot \phi_{j0}+2\sum\phi_{i0}\cdot \delta\phi_j
                        +\sum \delta\phi_i\cdot \delta\phi_j,
\end{equation}
where the mixed correlations $\phi_{i0}\cdot \delta\phi_j$ and  $\phi_{j0}\cdot \delta\phi_i$
are set identical when evaluated within the same scale range. The summation 
extends over $l_k\le r_{ij}\le l_{k+1}$ where $r_{ij}=\sqrt{(x_i-x_j) ^2+(y_i-y_j)^2}$ 
with coordinates $(x_i,y_i)$ for the position angle $\phi_i$.
The normalization factors are omitted here, and reintroduced later (therefore the 
sign '$\sim$').
Since observationally only $\phi_i$ is accessible and we wish to isolate 
$\sum \phi_{i0}\cdot \phi_{j0}$, we replace $\phi_{i0}=\phi_i-\delta\phi_i$
in the mixed correlation, which leads to:
\begin{equation}
\sum \phi_{i0}\cdot \phi_{j0} \sim \sum \phi_{i}\cdot \phi_{j}-2\sum\phi_{i}\cdot \delta\phi_j
                              +\sum \delta\phi_i\cdot \delta\phi_j.  \label{ac_expansion}
\end{equation}

In order to further proceed, the turbulent contribution $\delta\phi$ needs to be
quantified. Although $\delta\phi_i$ is not directly observable for an individual
position angle $\phi_i$, its distribution $\mathcal{D}(\delta\phi)$ can be constructed.
At the smallest observable scale $k=1$, the difference in two position angles,
$\Delta \phi_{ij}=\phi_i-\phi_j$, is mostly reflecting the difference in their turbulent
contributions. This is similar to the turbulent dispersion value which is 
derived by extrapolating from the smallest scales to the intercept at 
scale $k=0$ (section \ref{structure} and \citet{hildebrand09}), with the 
difference that $\Delta \phi_{ij}$ can be positive or negative. The 
turbulent distribution $\mathcal{D}(\delta \phi)$ can then be constructed 
as
\begin{equation}
\mathcal{D}(\delta \phi)=\frac{1}{2}\cdot\frac{b}{<\Delta \phi^2>^{1/2}_{k=1}}
                          \cdot\mathcal{D}(\Delta \phi_{k=1})
                           \label{turb_dist}
\end{equation}
where the distribution of $\Delta \phi_{k=1}$ is evaluated from the derived
differences in position angles at the smallest scale $k=1$, and the factor
$1/2$ results from assigning half of the difference as turbulent 
contribution to each of the two $PA$s.  The additional factor 
$b/<\Delta \phi^2>^{1/2}_{k=1}$ down-weights the measured differences
at scale $k=1$ to the limiting scale $k=0$, based on the result from the 
fit for the turbulent dispersion function (section \ref{structure}).

\begin{figure}
 \begin{center}
 \includegraphics[scale=0.7]{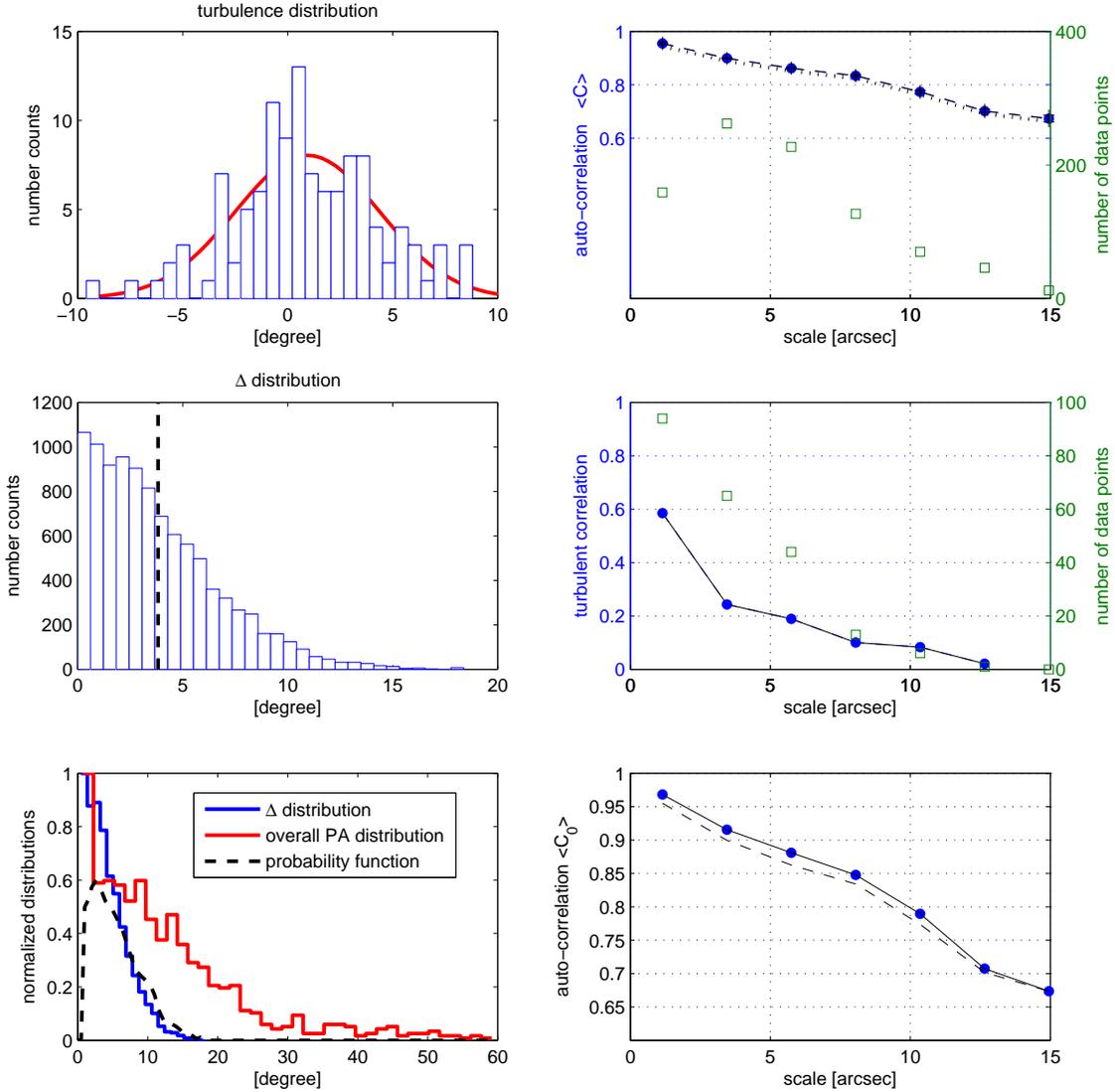}
 \caption{\label{corr_sep}\footnotesize
Illustration of the analysis of the polarization angle correlation function for W51, BIMA. 
Top left panel: the turbulence distribution $\mathcal{D}(\delta\phi)$, 
equation (\ref{turb_dist}), with mean $\mu\approx 0.9$. The resulting correction 
to the raw auto-correlation $<\mathcal{C}>$ -    
ideally identical to zero for a Gaussian turbulence distribution around zero -  
is shown in the top right panel with the dotted line. The dashed line shows
$<\mathcal{C}>$ calculated directly from the measured position angles $\phi_i$.
Middle left panel: the distribution of differences between two turbulence values, 
$\mathcal{D}(\Delta)$, derived by random sampling two values from the distribution
in the upper left panel. The larger number counts results from oversampling 
(calculating more difference pairs than combinations in  $\mathcal{D}(\delta\phi)$).
This also ensures a smooth distribution with a mean converging to $\mu_{\Delta}\approx 3.8$
(dashed line), which is used as a cut off criteria for  
the turbulent polarization angle correlation
function (middle right panel). A statistical 
weight (number counts) is applied (see appendix).
The large-scale auto-correlation $<\mathcal{C}_0>$ 
(bottom right panel, solid line) is obtained by adding
$<\mathcal{C}>$ and the 
turbulent correlation
with the proper normalization
from equation (\ref{ac_large_scale}). The dashed line shows again 
$<\mathcal{C}>$ for comparison. The only small correction from the 
turbulent correlation 
results from its small statistical weight
compared to the full distribution
of differences in $PA$s which is used for $<\mathcal{C}>$.
Bottom left panel: the normalized distributions of $\mathcal{D}(\Delta)$ (blue line)
and $\mathcal{D}(\Delta \phi)$ (red line) which are used to derive the probability
function (black dashed line) as an alternative to the cut off criteria 
in the middle left panel.
}
 \end{center}
\end{figure}

Typically, the turbulence distribution $\mathcal{D}(\delta \phi)$ is expected 
to be Gaussian around zero. This further means that for a large enough, 
statistically complete sample, the second term on the right hand side of 
equation (\ref{ac_expansion}) will vanish
due to the symmetry of $\mathcal{D}(\delta \phi)$. In order to verify this, the 
mixed term, $\sum\phi_{i}\cdot \delta\phi_j$, is calculated by randomly 
choosing a $ \delta\phi_j$ value from $\mathcal{D}(\delta \phi)$. Except for the 
smallest scale, there are fewer correlation products than sample values
$ \delta\phi_j$ in $\mathcal{D}(\delta \phi)$, and the mixed term will 
therefore typically not converge to zero. This statistical sampling 
limitation can be overcome by repeating the calculation and averaging the 
results. Typically, for 100 runs or more the results converge and then really
probe the statistical completeness of the turbulence distribution. 
Figure \ref{corr_sep} illustrates this for the case of W51 observed with BIMA: 
 $\mathcal{D}(\delta \phi)$, top left panel, is very close to a Gaussian 
distribution around
zero ($\mu\approx 0.9^{\circ}$). The resulting mixed term 
$\sum\phi_{i}\cdot \delta\phi_j$, the separation between the dotted and 
dashed line in the top right panel, is close to zero.

It remains to evaluate the third term on the right hand side of 
equation (\ref{ac_expansion}), $\sum\delta\phi_{i}\cdot \delta\phi_j$, 
which is the small scale turbulent correlation. Whereas the mixed term
extends over all scales, because it is a correction term due to a local
turbulent fluctuation with any possible position angle in each scale 
range, the turbulent correlation is expected to extend over a spatially limited 
area, characterized by the turbulent correlation length. In \citet{houde09} 
this is taken into account by assuming a Gaussian window function where the
width is the turbulent correlation length. In their further analysis the 
correlation length is then a fitting parameter. Here, we are directly
making use of the turbulence statistical distribution to set a cut off
in scale for the correlation term $\sum\delta\phi_{i}\cdot \delta\phi_j$.
As in the case of  $\sum\phi_{i}\cdot \phi_j$, the relevant quantity
for the correlation is the difference, $\Delta=|\delta \phi_i - \delta \phi_j|$,
between a pair of turbulence fluctuations $\delta \phi_i$ and $\delta \phi_j$.
The distribution of $\Delta$ can again be constructed by calculating 
differences between two randomly chosen values from the distribution
$\mathcal{D}(\delta \phi)$. $\mathcal{D}(\Delta)$ is shown for W51 (BIMA)
in Figure \ref{corr_sep}, middle left panel, 
with a resulting mean difference in turbulence fluctuations
$\mu_{\Delta}\approx 3.8^{\circ}$. We are adopting $\mu_{\Delta}$ as 
a cut off measure. This means that  $\delta\phi_{i}\cdot \delta\phi_j$ 
is counted as a correlation pair for a a certain scale $k$ if the 
difference in the corresponding $\phi_i \cdot \phi_j$ pair, 
$\Delta \phi=|\phi_i - \phi_j|$, is smaller than $\mu_{\Delta}$. 
Otherwise the correlation in $\phi_i \cdot \phi_j$ is considered to be
dominated by the large-scale magnetic field. 
Alternatively, the cut off criteria can be refined by calculating the 
probability function of $\Delta \phi$ being of turbulent origin: to that
purpose, the normalized distributions of $\mathcal{D}(\Delta)$ and
$\mathcal{D}(\Delta \phi)$ are weighted against each other, Figure
\ref{corr_sep}, bottom left panel. (When working with the normalized
distributions, the weighting factors (see below) are not yet taken 
into account.)
Comparing the distribution
$\mathcal{D}(\Delta)$ with the distribution $\mathcal{D}(\Delta \phi)$
provides thus a tool to separate small and large-scale contributions. 
Both the $\mu_{\Delta}$ cut off and the more sophisticated weighting
with the probability function lead to very similar results. We therefore
adopt the simpler criteria with $\mu_{\Delta}$ in the following. 
Although $\mathcal{D}(\Delta)$ is only known in a statistical way, 
with the same $\mu_{\Delta}$ used for each correlation pair, this 
average value can still be used to check against the measured 
difference in each correlation pair $\phi_i \cdot \phi_j$. Technically, 
we thus evaluate:
\begin{equation}
\delta\phi_{i}\cdot \delta\phi_j = \phi_{i}\cdot \phi_j  \hspace{1cm} {\rm if}
                                   \hspace{1cm}\Delta\phi \le \mu_{\Delta}
                                   \label{turb_condition}
\end{equation} 
When adding the two correlation parts in equation (\ref{ac_expansion}), 
 $\sum\phi_{i}\cdot \phi_j$ and  $\sum\delta\phi_{i}\cdot \delta\phi_j$,
a proper normalization needs to be introduced. A factor $1/(N_k+N_{\delta,k})$
with $N_{\delta,k}$ counting the correlation pairs satisfying 
equation (\ref{turb_condition}), normalizes the sum to one.
The large-scale correlation function, $<\mathcal{C}_0>$, for each 
scale $k$ can then be written as
\begin{equation}
<\mathcal{C}_0>_k=\frac{1}{N_k+N_{\delta,k}}\left(
\sum_k\phi_{i}\cdot \phi_j+\sum_k\delta\phi_{i}\cdot \delta\phi_j\right)
                                  \label{ac_large_scale}
\end{equation}
with the condition in equation (\ref{turb_condition}).

The isolated 
turbulent correlation, 
$1/N_{\delta,k}\cdot \sum\delta\phi_{i}\cdot \delta\phi_j$, is shown 
in Figure \ref{corr_sep}, middle right panel. A weighting factor
$N_{\delta,k}/N_k$ is taken into account to directly compare its
statistical significance with $<\mathcal{C}>_k$ in the right upper most panel
(weighting factor $N_k/N_k\equiv 1)$.
This additional weighting provides a spatial filter similar to assuming
a Gaussian window function \citep{houde09}. It acts like a probability, 
for each scale $k$, that a measured $\Delta \phi \le \mu_{\Delta}$ 
contributes to the turbulent correlation. We note that this is also 
valid when adopting the probability function as a cut off criteria, 
as mentioned above.

The bottom right panel in Figure \ref{corr_sep} shows the final 
large-scale correlation function
$<\mathcal{C}_0>_k$ by combining the upper two panels with the proper
normalization from equation (\ref{ac_large_scale}). Although the 
turbulent correlation 
shows a relatively sharp decrease over the smallest scales, its 
overall correction to $<\mathcal{C}>_k$ (dashed line) is small. This is due 
to the small statistical significance (in terms of number counts) of the 
isolated turbulence differences in $PA$s. 
Adding up the various correlation factors with their statistical weight is
one possible choice. Its most immediate advantage is that it gives 
directly the correct normalization.

Finally, we note that the characteristic large-scale 
polarization angle correlation length, 
$\lambda_{0}$, is thus virtually unchanged
compared to when it is directly calculated from 
the measured ensemble of $\phi_i$.
From the middle panel, 
additionally, a turbulent polarization angle 
correlation length, $\lambda_{t}$, can be estimated 
in an analog manner.



\begin{deluxetable}{ccccccccccccccc}
 \tabletypesize{\scriptsize}
\tablewidth{0pt}
\tablecaption{Observation Summary and Results from the Statistical
Analysis
                      \label{quantities}}
\tablehead{
\multicolumn{5}{c}{Observation}
             & \colhead{}
             & \multicolumn{9}{c}{Analysis}\\
  \cline{1-6} \cline{8-15}
\colhead{instrument} & \colhead{$\lambda$}  & \colhead{$\theta$}  &
\colhead{$\theta$}  & \colhead{$d$} &
\colhead{ref} &  \colhead{} &
\colhead{$b$} &  \colhead{$\frac{<B_t^2>^{1/2}}{B_0}$} & \colhead{$m$} &
\colhead{$\alpha$}  &
\colhead{$\lambda_{0}$} & \colhead{$\lambda_{t}$} & \colhead{$N$}  
& \colhead{$\left(\frac{<B_t^2>^{1/2}}{B_0}\right)_N$}\\
 \colhead{} &  \colhead{(mm)} &  \colhead{($\arcsec$)} & \colhead{(mpc)} &
\colhead{(mpc)} & \colhead{} & \colhead{}
& \colhead{(deg)} & \colhead{}  & \colhead{(deg/$\arcsec$)} & \colhead{}
& \colhead{(mpc)} & \colhead{(mpc)} & \colhead{} & \colhead{}\\
 \colhead{} &  \colhead{(1)} &  \colhead{(2)} &  \colhead{(3)} &  \colhead{(4)} &  \colhead{(5)} &
 \colhead{} &  \colhead{(6)} &  \colhead{(7)} &  \colhead{(8)} &  \colhead{(9)} & \colhead{(10)}
 & \colhead{(11)}  & \colhead{(12)}  & \colhead{(13)} 
}
\startdata
w51& & & & & & & & &&\\
\cline{1-1}\\
BIMA       &  1.3     &  2.3      &  69    & 300 &   I &  &  6.4        &
0.08  & 2  & 1.1   &  230  &  45 & 122$^d$ & 1.22 \\
SMA - e2   &  0.87    &  0.7    &   21    & 60  &   II &  &  27.6$^c$   &
0.36  & 35 & 1.3   &   73  &  25 & 4$^d$  & 0.70\\
SMA - e8   &  0.87    &  0.7    &   21    & 60  &   II &  &  54.0$^c$   &
0.89  & 10 & 0.3   &   63  &  30 & 4$^d$  & 1.27\\
 & & & & & & & & &\\

Orion  & & & & & & & &  & & & && \\
\cline{1-1}\\

BIMA      &  3       &   7.0     &  16.1    & 69  &   III &  &  14.6    &  0.18
& 3  & 1.6   &  66  &  36  & 6 & 0.44\\
BIMA      &  1.3     &   3.4     &     7.8  & 69  &   III &  &  12.9   &  0.16
& 10  & 3.2   &  46  &  -  & 4  & 0.30\\
SMA$^a$   &  0.87    &   2.8     &     6.4  & 23  &   IV &  &  24.7   &  0.32
& 7  & 1.3   &  35  &  10  & 1 & 0.33\\
SMA$^b$   &  0.87    &   0.9     &     2.1  & 23  &   IV &  &  33.1   &  0.45
& 10 & 0.7   &  15  &  9  & 1 & 0.43\\

\enddata
\tablecomments{All statistical quantities are obtained from
the Figures \ref{w51_stat} and \ref{orion_stat} with the method described
in section \ref{method}.
}
\tablenotetext{(1)}{observing wavelength}
\tablenotetext{(2)}{synthesized beam resolution, also used as the separation
in between the bins $l_k$; for elliptical beams the geometrical mean is adopted, 
except for Orion BIMA 3mm where the semi-major axis is used because of its
 very elliptical beam}  
\tablenotetext{(3)}{synthesized beam in physical scale at source distance}
\tablenotetext{(4)}{scale of variation in the large-scale polarization structure,
empirically estimated from polarization maps by visual inspection}
\tablenotetext{(5)}{references: (I) \citet{lai01}, (II) \citet{tang09b}, (III) \citet{rao98},
(IV) \citet{tang09c}}
\tablenotetext{(6)}{turbulent magnetic field dispersion}
\tablenotetext{(7)}{turbulent to mean field strength ratio}
\tablenotetext{(8)}{slope over the first two bins (W51) or three bins
(Orion) in the dispersion function $<\Delta \phi ^2>^{1/2}$}
\tablenotetext{(9)}{power law index of the second order structure function,
                  $<\Delta \phi ^2> \sim l^{\alpha}$, over the first two
bins (W51) or three bins (Orion)}
\tablenotetext{(10)}{plane of sky projected large-scale polarization angle correlation length}
\tablenotetext{(11)}{plane of sky projected turbulent polarization angle 
                     correlation length, derived from the method in Appendix A}
\tablenotetext{(12)}{number of turbulent cells contained within telescope beam}
\tablenotetext{(13)}{turbulent to mean field ratio corrected for beam integration effect}
\tablenotetext{a}{compact and subcompact observation combined}
\tablenotetext{b}{compact and extended observation combined}
\tablenotetext{c}{possibly influenced by gravity, with a bias of $\sim -7$
deg (section \ref{validity})}
\tablenotetext{d}{based on 
approximating  $\Delta^{\prime}$ (effective depth of the molecular cloud along the line of sight)
with the size of emission which is roughly the size of the maps in the SMA and BIMA observations.
The turbulent correlation length $\delta$ is adopted from \citet{houde09}.  
}
\end{deluxetable}

\begin{deluxetable}{c|ccc}
\tablewidth{0pt}
\tablecaption{Power Law Indices $\gamma$ for Scaling Relations
                      \label{power}}
\tablehead{
\colhead{parameter} & \colhead{W51}  & \colhead{Orion$^a$}  &
\colhead{combined$^b$}
}

\startdata

$<B_t ^2>^{1/2}/B_0$ (uncorrected)  &  -1.51   &  -0.43   & -0.33 \\
$(<B_t ^2>^{1/2}/B_0)_N$ (beam corrected)  &  0.17   &  0.07   & -- \\
$\lambda_{0}$              &  1.02    &  0.76    &  0.75  \\

\enddata
\tablecomments{All power law indices $\gamma$ are derived by fitting the
relation $\propto l^{\gamma}$
as a function of physical scale $l$ (Figure \ref{scaling_loglin}).
Separate fits are performed to subsets (W51, Orion) and the entire data
set.
}
\tablenotetext{a}{includes Orion from \citet{hildebrand09} for 
$<B_t ^2>^{1/2}/B_0$ (uncorrected and beam corrected)}
\tablenotetext{b}{W51 and Orion combined. Orion, M17 and DR21  from \citet{hildebrand09}
are included for $<B_t ^2>^{1/2}/B_0$ (uncorrected). No joint fit is attempted with 
the beam corrected data.
$\lambda_{0}$ is fit from the combined SMA and BIMA data, analyzed in the work here.}

\end{deluxetable}

\end{document}